\documentclass[aps,pre,twocolumn,superscriptaddress,floatfix,amsmath]{revtex4}
\usepackage{graphicx}
\usepackage{psfrag}
\usepackage{bm}
\usepackage{wasysym}
\usepackage{units}

\begin{document}
\title{Experimental test of curvature-driven dynamics
in the phase ordering of a two dimensional liquid crystal}
\author{Alberto Sicilia}
\affiliation{Universit\'e Pierre et Marie Curie -- Paris VI, LPTHE UMR 7589,
4 Place Jussieu,  75252 Paris Cedex 05, France}
\author{Jeferson J. Arenzon}
\affiliation{Instituto de F\'\i sica, Universidade Federal do
Rio Grande do Sul, CP 15051, 91501-970 Porto Alegre RS, Brazil}
\author{Ingo Dierking}
\affiliation{School of Physics and Astronomy, University of Manchester,
Manchester M13 9PL, UK}
\author{Alan J. Bray}
\affiliation{School of Physics and Astronomy, University of Manchester,
Manchester M13 9PL, UK}
\author{Leticia F. Cugliandolo}
\affiliation{Universit\'e Pierre et Marie Curie -- Paris VI, LPTHE UMR 7589,
4 Place Jussieu,  75252 Paris Cedex 05, France}
\author{Josu Mart\'inez-Perdiguero}
\affiliation{Departamento de F\'isica de la Materia Condensada, Facultad 
de Ciencias, Universidad del Pa\'{\i}s Vasco, Apartado 644, 48080 Bilbao, Spain}
\author{Ibon Alonso}
\affiliation{Departamento de F\'isica de la Materia Condensada, Facultad 
de Ciencias, Universidad del Pa\'{\i}s Vasco, Apartado 644, 48080 Bilbao, Spain}
\author{Inmaculada C. Pintre}
\affiliation{Qu\'{\i}mica Org\'anica, Facultad de Ciencias, Instituto de Ciencia 
de Materiales de Arag\'on, Universidad de Zaragoza-CSIC, 50009 Zaragoza, Spain}

\date{\today}  
\begin{abstract}
  We study electric field driven deracemization in an achiral liquid crystal
  through the formation and coarsening of chiral domains.
  It is proposed that deracemization in this system
  is a curvature-driven process. We test this prediction using the
  exact result for the distribution of hull-enclosed areas in
  two-dimensional coarsening in non-conserved scalar order parameter dynamics
  recently obtained [J.J. Arenzon et al., Phys.\ Rev.\ Lett.\ {\bf 98}, 061116
  (2007)]. The experimental data are in very good agreement with the
  theory. We thus demonstrate that deracemization in such bent-core liquid
  crystals belongs to the Allen-Cahn universality class, and
  that the exact formula, which gives us the statistics of domain sizes
  during coarsening, can also be used as a strict test for this dynamic
  universality class.
\end{abstract}
\maketitle

Many aspects of the nonequilibrium relaxation of macroscopic systems
still remain to be grasped. The domain growth of two competing
equilibrium phases after a quench from the disordered phase is a
relatively simple out of equilibrium problem and in some cases the
mechanism underlying coarsening is well understood (curvature driven,
bulk diffusion, etc.)~\cite{BrayReview}. However, an important part of
the description of these processes remains phenomenological and, to a
certain extent, qualitative.

According to the scaling hypothesis, a system in the late stages 
of coarsening is described by a scaling phenomenology in which 
there is a single characteristic length scale (``domain scale''), 
$R(t)$, that grows with time. In consequence, all
dynamical properties occuring on scales large compared to 
microscopic ones are described by scaling functions in which lengths 
are scaled by $R(t)$. For example the pair correlation function, 
$C(r,t) = \langle S({\bf x},t)\,S({\bf x} + {\bf r},t) \rangle$, where 
$S$ takes the values $\pm 1$ in the two equilibrium phases, has the 
scaling form $C(r,t) = g[r/R(t)]$ \cite{BrayReview}. Verifying the 
scaling hypothesis, and computing the scaling functions, has been a 
longstanding challenge. Recently, however, significant progress was made 
when the distribution $n_h(A,t)$ of hull enclosed areas (those enclosed by the outer 
boundaries of domains) was computed for scalar non-conserved
order parameter dynamics in two dimensions and the scaling hypothesis 
verified for this quantity \cite{us-PRL,us-PRE}. The analytically obtained 
distribution function was shown to be robust and hold -- to a high numerical
precision -- in Monte Carlo simulations of the pure~\cite{us-PRL,us-PRE} 
and disordered~\cite{us-EPL} bidimensional kinetic Ising Model ($2d$IM).
The question remains as to whether more complicated experimental 
systems could also be described by such a universal formula.

In this Letter we test this result experimentally in a liquid crystal
system and we find a very good agreement with the theory.  We thus
demonstrate that the system belongs to the universality class of
non-conserved scalar order parameter dynamics and that the exact formula 
is a universal property of these systems.  In the following
paragraphs we describe the experiment, and present a detailed analysis
of the data.


The experimental system chosen is one that exhibits electric field
driven deracemization. Since the discovery of Louis Pasteur, more than
150 years ago, that chiral crystals can form from an achiral solution
\cite{Pasteur}, deracemization has been a fundamental question in the
investigation of chirality.  More recently, practical applications in,
for example, drug design and synthesis, boosted research in this
field, as the effect of most modern drugs is based on chiral
molecules.  
Spontaneous  deracemization in an  achiral fluid  system 
is   very  unusual   and  a  topic   of  only   recent  interest
\cite{Takanishi}.    It  can  occasionally   be  observed   in  liquid
crystalline  systems  \cite{Thisayukta,Heppke,Rao},  mainly formed  by
bent-core or  so called ``banana''  molecules. The most  likely reason
for chiral conglomerate formation is steric interactions. This is also
evidenced by  computer simulations  and theory, which  indicate chiral
conformations     of    on     the    average     achiral    molecules
\cite{Earl}. Electric  field induced switching  between chiral domains
was demonstrated  in Ref.~\cite{Eremin}. Kane  {\it et al.}~\cite{Kane} very
recently exhibited  the electric field driven  deracemization of an
achiral  fluid  liquid crystalline  system,  and  gave a  theoretical
interpretation of the conglomerate  formation in terms of a difference
in the  chemical potential  of left and  right handed  molecules under
electric field application.

The liquid crystal employed in this investigation is comprised of a
bent-core molecule, which together with cell preparation conditions is
discussed in detail in Ref.~\cite{Martinez}. The studied cell has a
gap of 5$\mu{\rm m}$ filled with the liquid crystal, while lateral
dimensions are much larger, approximately 1cm in each direction. We
are thus effectively investigating a two-dimensional system. Domain
coarsening was followed by temperature controlled polarising
microscopy (Nikon Optiphot-Pol microscope in combination with a
Linkham TMS91 hot stage), with a control of relative temperatures to
0.1K. Digital images were captured at a time resolution of 1s with a
pixel resolution of $N=1280\times 960$, corresponding to a sample size of
$520 \times 390 \mu{\rm m}^2$ (JVC KY-F1030). Note that the imaging
box is approximately 1/200 of the whole sample.
Electric square-wave fields of amplitude $E=14V\mu{\rm m}^{-1}$ and
frequency $f = 110{\rm Hz}$ were applied by a TTi-TG1010 function
generator in combination with an in-house built linear high voltage
amplifier.

Cooling from the isotropic liquid, an optically isotropic fluid liquid
crystal phase is formed, which exhibits no birefringence and thus
appears dark between crossed polarisers.  The phase transition is
first-order.  On electric field application, chiral deracemization
occurs with domains of opposite handedness growing as a function of
time. This coarsening process can easily be followed when the
polarisers are slightly de-crossed by a few degrees. The chiral
domains of opposite handedness get larger as smaller domains
disappear. At the same time the area distribution of domains with
opposite handedness remains constant at an equal distribution of left-
and right-handed domains, because there is an overall constraint of
zero chirality over the full sample that needs to be respected.  This
was checked experimentally and found to be true within experimental
uncertainty due to finite size of the image window and thresholding. Still,
such a global constraint (as opposed to a local one) is not expected
to change the coarsening universality class which remains
curvature-driven~\cite{BrayReview}.

\begin{figure}[h]




\includegraphics[width=3.5cm,angle=90]{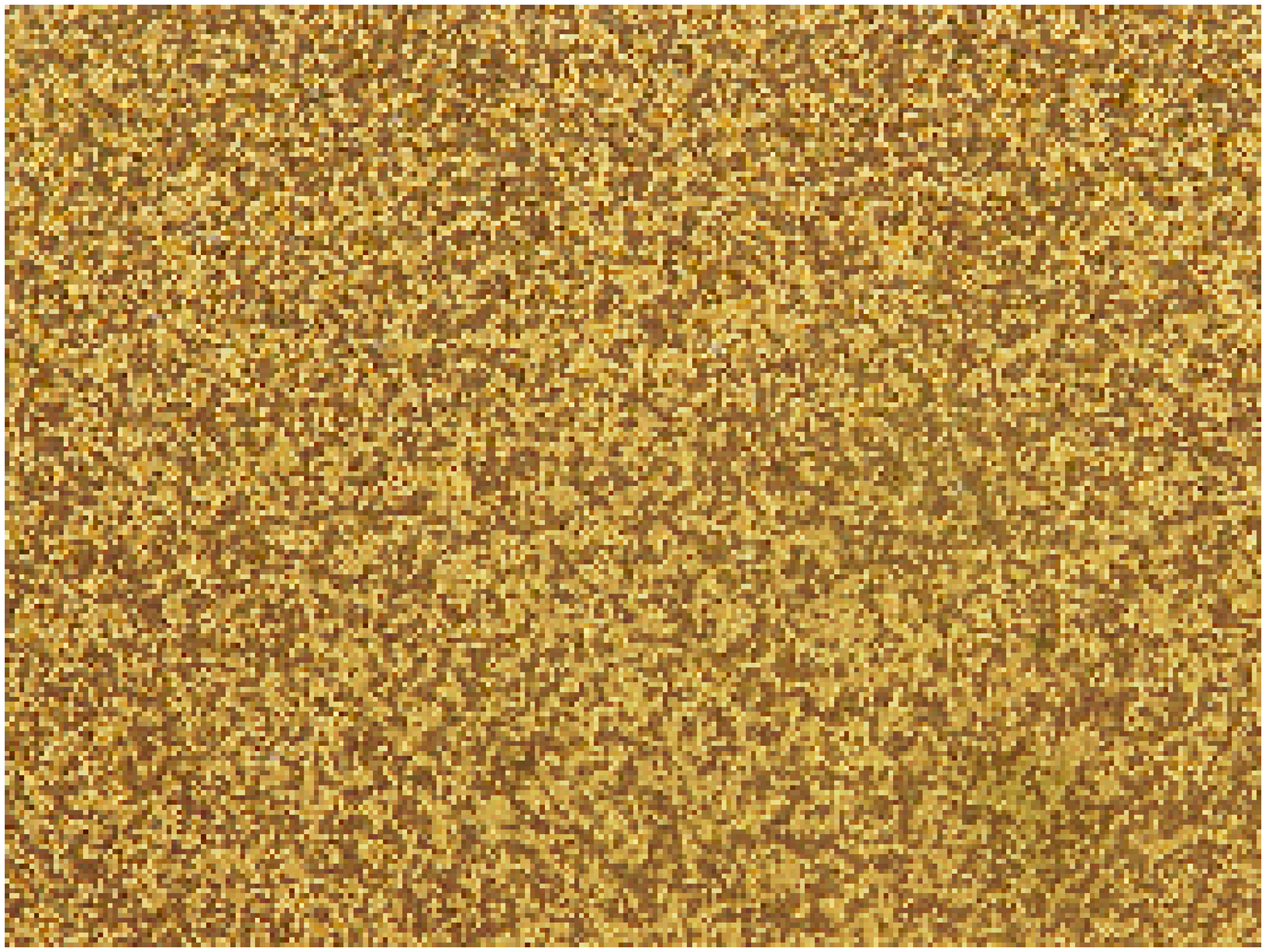}
\includegraphics[width=3.5cm,angle=90]{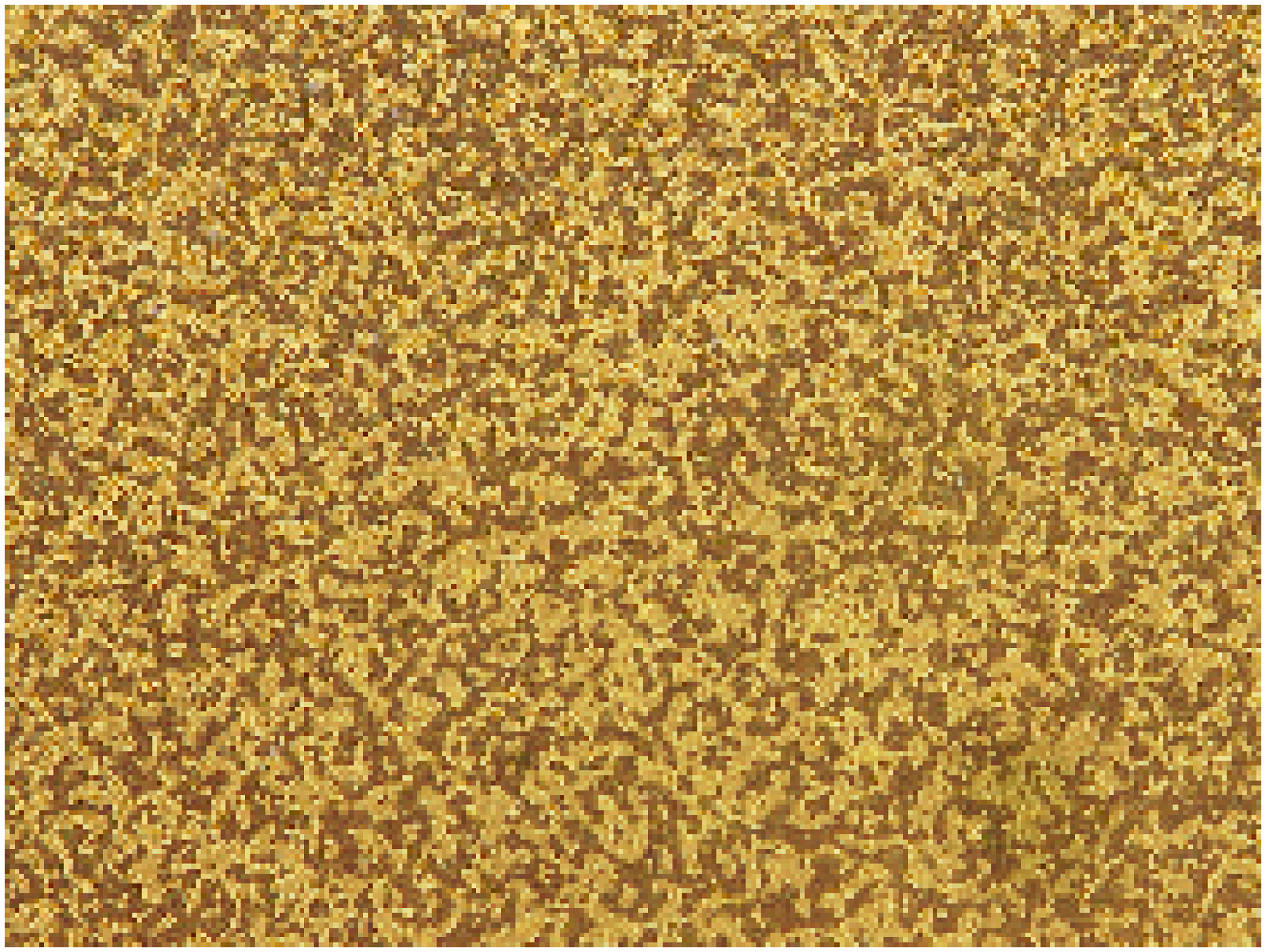}
\includegraphics[width=3.5cm,angle=90]{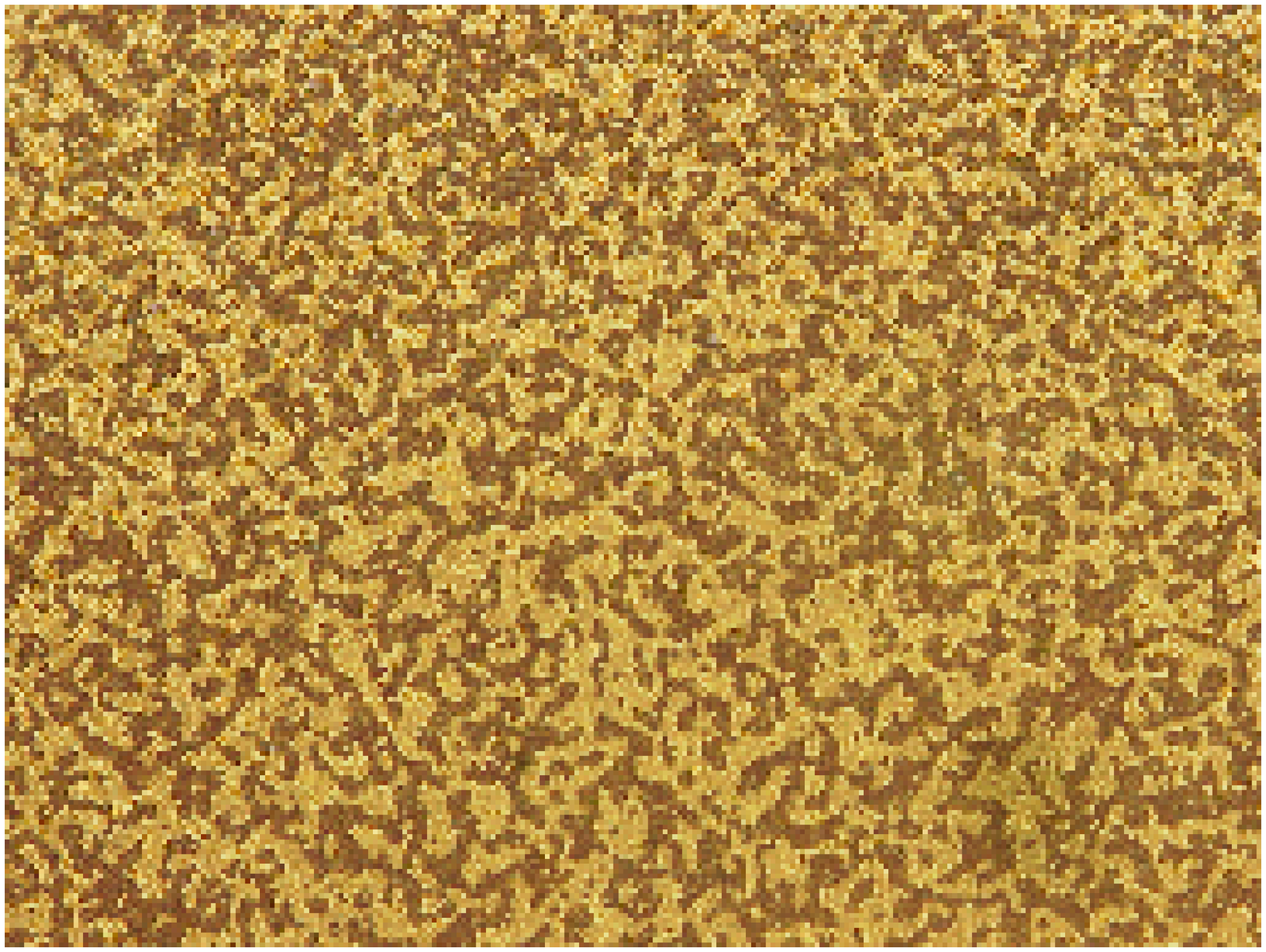}

\vspace{0.12cm}

\includegraphics[width=3.5cm,angle=90]{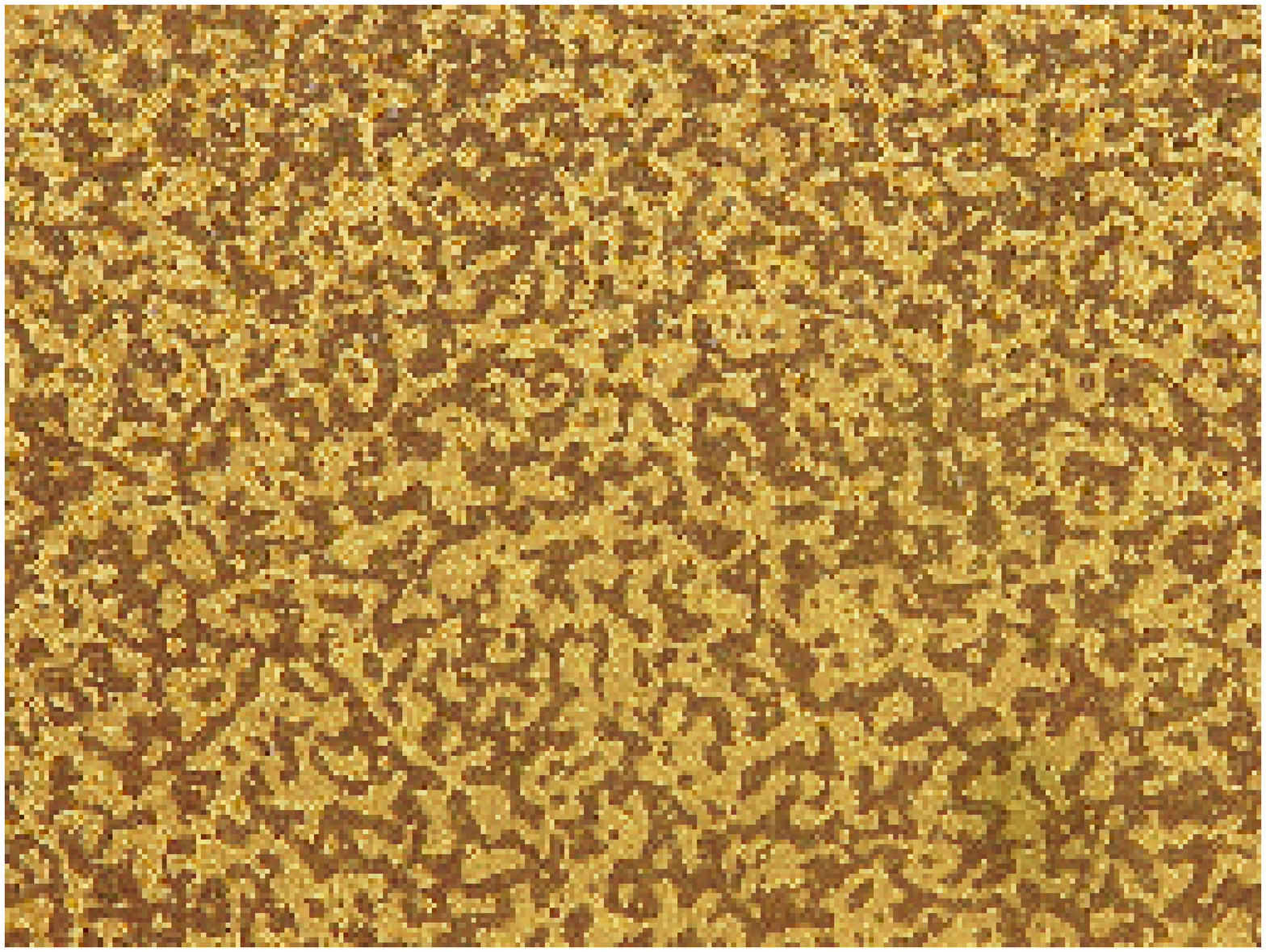}
\includegraphics[width=3.5cm,angle=90]{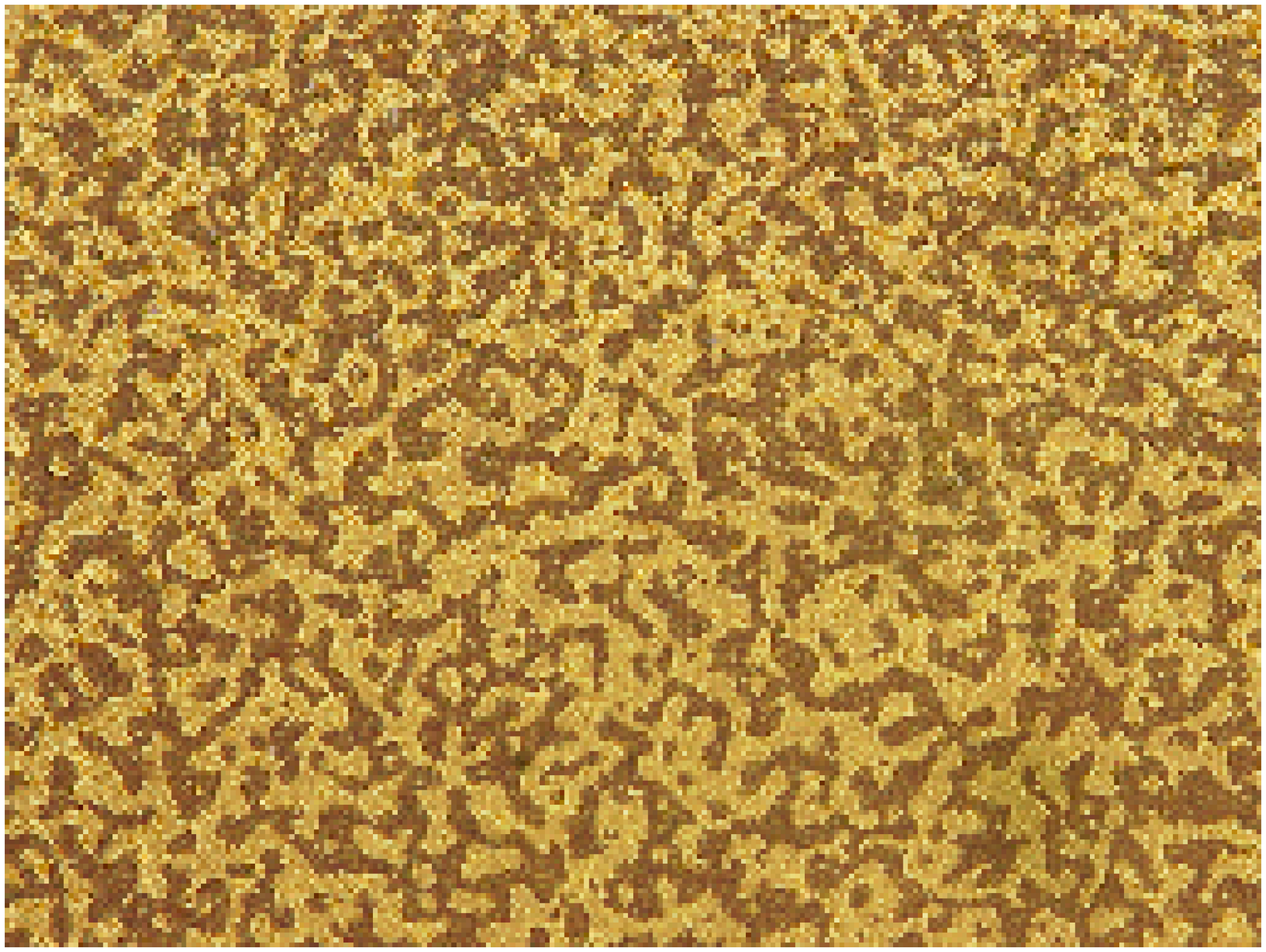}
\includegraphics[width=3.5cm,angle=90]{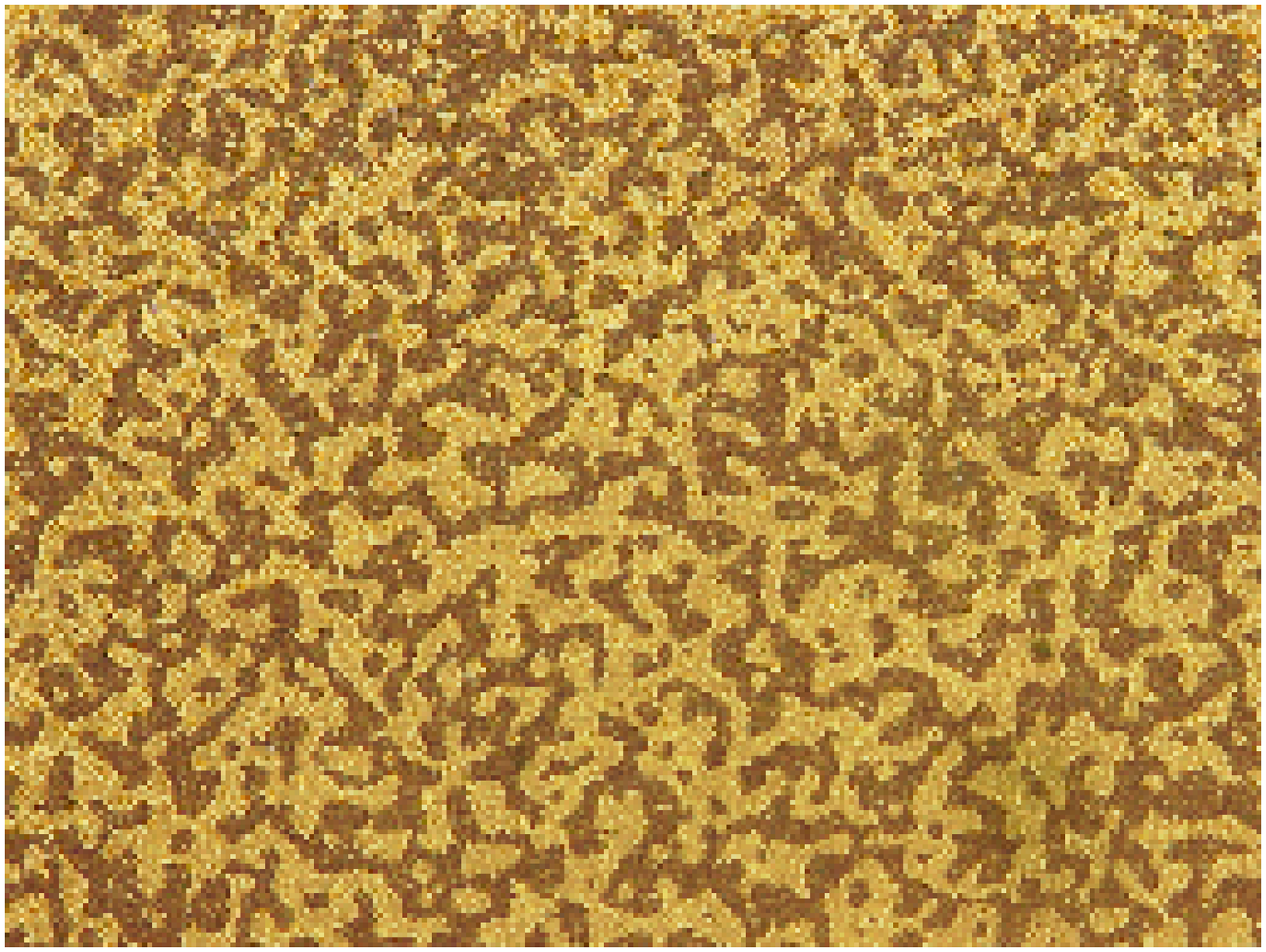}
\caption{The first snapshot displays the configuration right after the quench,
$t=\unit[0]{s}$. The others are snapshots during the evolution $t=60,120,...,\unit[300]{s}$.}
\label{fig.snapshots}
\end{figure}

We performed 10 runs lasting \unit[10]{min} each with pictures taken
at intervals of \unit[10]{s} on a single sample. Each run is
initialized by heating the sample above the transition
temperature and subsequently cooling below it. The coarsening process in the
low temperature phase is visualized in terms of domains, i.e., connected
regions of the same handedness.  In Fig.~\ref{fig.snapshots} we show a
series of snapshots taken at times $t=0, 60, 120,...,\unit[300]{s}$.
These pictures are then thresholded and an Ising
spin $s_i$ is assigned to each pixel, where $s_i(t) = \pm 1$ for pixels 
that belong to left or right handed
domains, respectively. There are many spurious
small domains that are related to the experimental system rather
than to thermal fluctuations. The induced graininess
is also reflected in the small $r$ behavior of the pair correlation
function $C(r,t)$ and the small $A$ behaviour of $n_h(A,t)$ as we shall 
see below. Still, at face value the domain
geometry is the one of scalar non-conserved order parameter dynamics,
as can be checked by comparing to the snapshots shown in \cite{us-PRE}
for the $2d$IM.

The initial ``magnetization density'' in the imaging window
of the liquid crystal, defined as the average of the spin variables
over the box, $m(0)=N^{-1} \sum_{i=1}^N s_i(0)$, is not zero.  This
initial value is only approximately conserved by the dynamics,
$m(t)\approx m(0)= 0.2 \pm 0.1$, but the actual value depends on the
thresholding operation.

\begin{figure}[h]
\includegraphics[width=6.9cm]{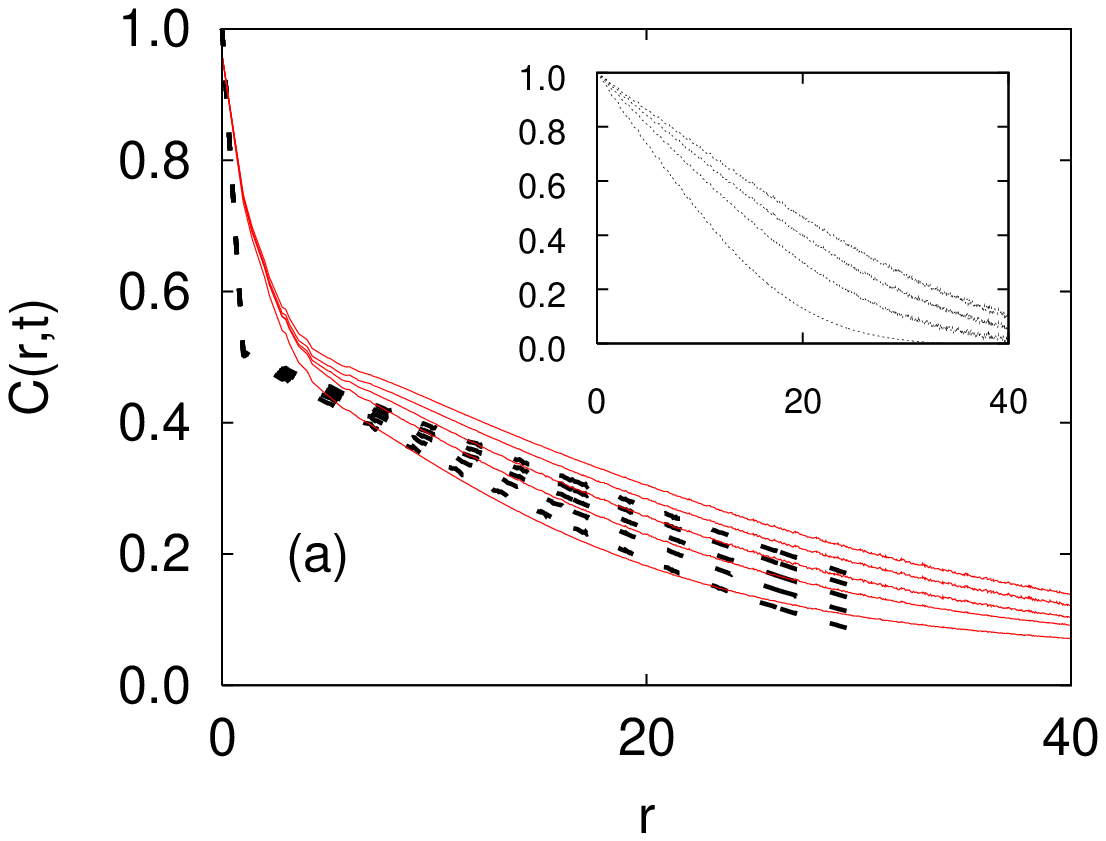}
\includegraphics[width=6.9cm]{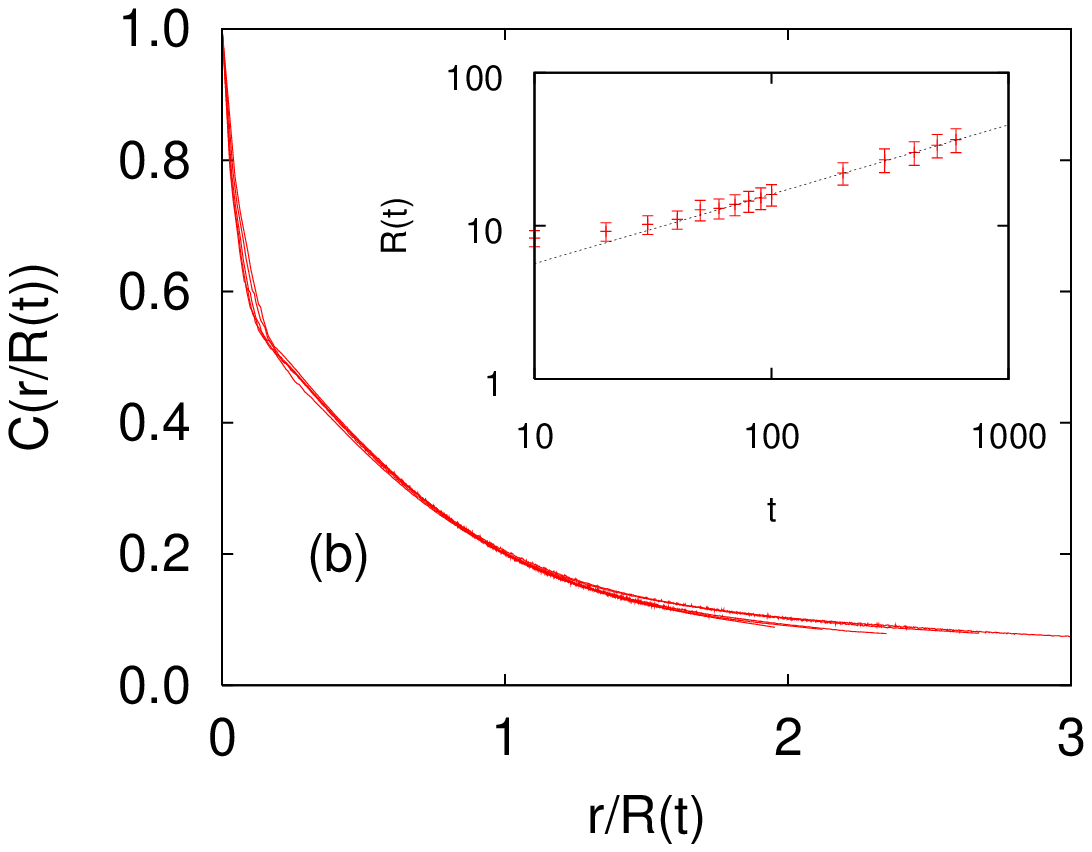}
\caption{(Colour online.)  Spatial correlation function at different
  times after the quench. (a) Experimental data at five equally spaced
  times, $t=100, \dots, 500$s with thin (red) lines, and numerical
  simulation data in the $2d$IM with $10\%$ randomly flipped spins at
  five equally spaced times, $t=100, \dots, 500$ MCs with dashed
  (black) lines.  We clearly notice the effect of graininess at very
  small scales ($r\stackrel{<}{\sim}5$ in the experiment and
  $r\stackrel{<}{\sim} 1$ in the simulation), where there is no
  time-dependence in either case. Inset: the actual spatial
  correlation in the $2d$IM.  (b) Study of the scaling hypothesis,
  $C(r,t) \simeq g[r/R(t)]$ in the liquid crystal, at the same times
  as in panel (a). Inset: the time-dependence of the growing-length
  scale. The slope of this line is $1/z\simeq 0.45\pm 0.10$.}
\label{fig.correlation}
\end{figure}

We determine the growth law for the size, $R(t)$, of typical domains 
from a direct measure of the spatial correlation function,
$
C(r,t) \equiv \frac{1}{N} \sum_{i=1}^N
\langle \, s_i(t) s_j(t) \, \rangle
\Large{|}_{|\vec r_i - \vec r_j|=r}
$,
The angular brackets indicates an
average over the 10 runs.  The distance dependence of the
pair-correlation at five equally spaced times, $t=100, \dots, 500$s is
displayed with thin (red) lines in Fig.~\ref{fig.correlation}(a). As a
consequence of the non-zero magnetization, $C(r,t)$ does not decay to
zero at large $r$.  More strickingly, the curves are time-independent
at distances $r\stackrel{<}{\sim} 5$ ($C\stackrel{>}{\sim} 0.55$) and
they clearly depend on time at longer distances with a slower decay
at longer times. Here and in what follows we measure distances in
units of the lattice spacing. The time-independence at short-distances
and the long-distance decay are atypical, as can be seen by comparing
to the spatial correlation in the $2d$IM displayed in the inset to
Fig.~\ref{fig.correlation}(a). We ascribe the lack of time-dependence at
short scales and the further slow decay to the graininess of the experimental system.
Indeed, in Fig.~\ref{fig.correlation}(a) we also
show with thick dashed (black) lines the correlation in the $2d$IM
where we have flipped, at each measuring instant, $10\%$ of spins
taken at random over the sample (the system dynamics are not perturbed
and between measurements we use the original spins).  By comparing the
two sets of curves we see that the effect of the random spins is
similar to the one introduced by the graininess of the system.
This effect will also be important for the analysis of the hull-enclosed
area distribution.

The function $C(r,t)$ obeys dynamical scaling,
$C(r,t)
\simeq  g[r/R(t)]$.
We define the characteristic length-scale $R(t)$ at time $t$
by the condition $C(R,t)=0.2$ but other choices give equivalent
results. The good quality of the scaling is shown in
Fig.~\ref{fig.correlation}(b).  The time-dependence of the growing
length $R(t)$ is shown in the inset to Fig.~\ref{fig.correlation}(b)
with points.  The errorbars are estimated from the variance of the
values obtained from the 10 independent runs. We measure the growth
exponent $1/z$ by fitting the long-time behavior of $R(t)$, say for
$t>30$ s, and we find $1/z\simeq 0.45\pm 0.10$. The exponent thus
obtained is close to the theoretically expected value $1/2$ for clean
non-conserved order parameter dynamics~\cite{BrayReview}. The data suggest that
for times longer than $t\simeq 30$ s the system is well in the scaling
regime.

We now turn to the analysis of the distribution of hull enclosed
areas.  Each domain has one external perimeter which is called the
{\it hull}. The {\it hull-enclosed area} is the total
area contained within this perimeter. In~\cite{us-PRL,us-PRE} we
derived an exact analytical expression for the hull enclosed area distribution of
curvature driven two-dimensional coarsening with non-conserved order
parameter.  Using a continuum description in which the non-conserved
order parameter is a scalar field we found that the number of
hull-enclosed areas per unit area, $n_h(A,t)\,dA$, with enclosed area
in the interval $[A,A+dA]$, after a quench from high temperatures is
\begin{equation}
n_h(A,t)  =  
2c_h/(A+\lambda_h t)^2
\; .
\label{eq.analytic-nh}
\end{equation}
$c_h=1/8\pi\sqrt{3}$ is a universal constant that enters this
expression through the influence of the initial condition and was
computed by Cardy and Ziff in their study of the geometry of critical
structures in equilibrium~\cite{Cardy}.  $\lambda_h$ is a material
dependent constant relating the local velocity $v$ of an interface and its
local curvature $\kappa$, in the Allen-Cahn equation, $v=-(\lambda_h/2\pi) \,
\kappa$~\cite{AC}. 
Equation~(\ref{eq.analytic-nh}) can be recast in the scaling form
$n_h(A,t) = (\lambda_h t)^{-2} \; \;
f\left( A/\lambda_h t \right)
$,
with $f(x)=2c_h/(x+1)^2$.  In this way, scaling with the
characteristic length scale, $R(t) = \sqrt{\lambda_h t}$, for
coarsening dynamics with scalar non conserved order parameter in a pure
system is demonstrated. The effects of a finite working
temperature are fully encoded in the temperature dependence of 
$\lambda_h$ while the same scaling function $f(x)$ describes
$n_h$~\cite{us-PRE} as suggested by the zero temperature fixed point
scenario~\cite{BrayReview}.

\begin{figure}[h]
\includegraphics[width=6cm]{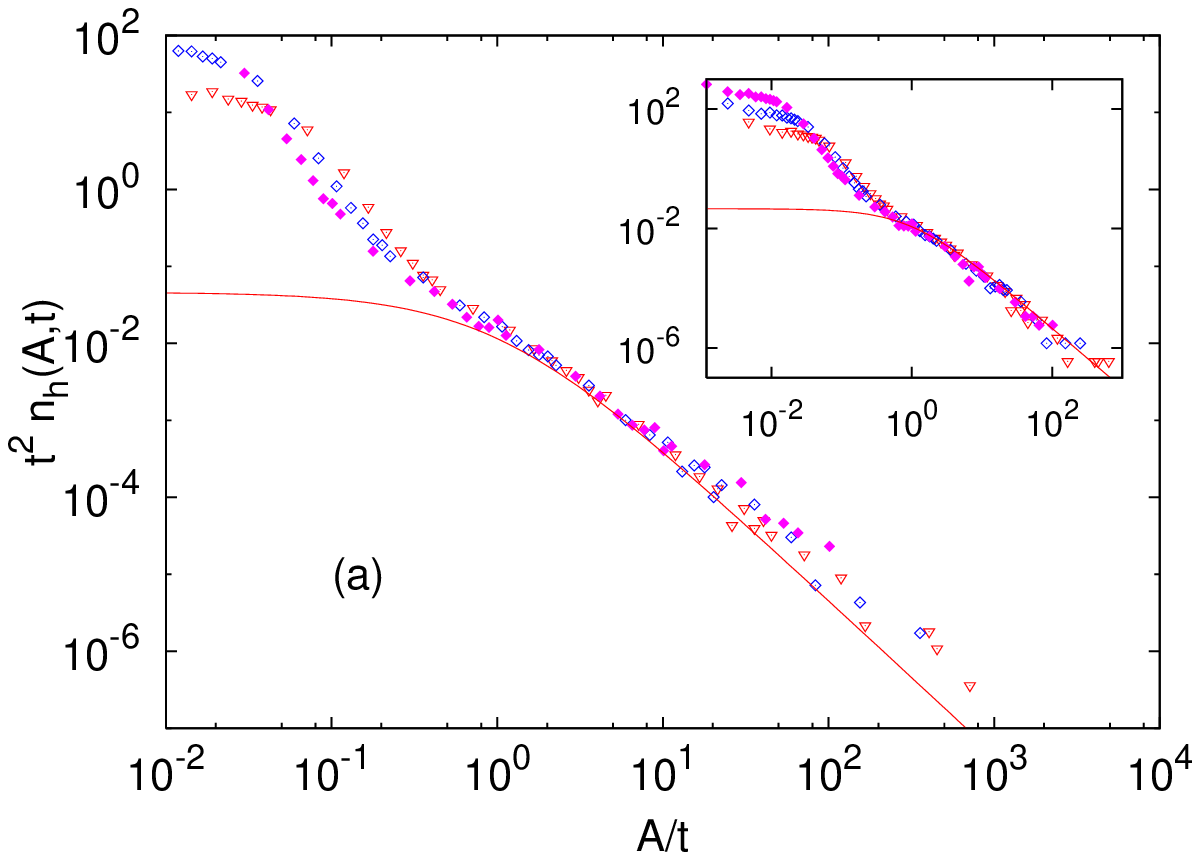}
\includegraphics[width=6cm]{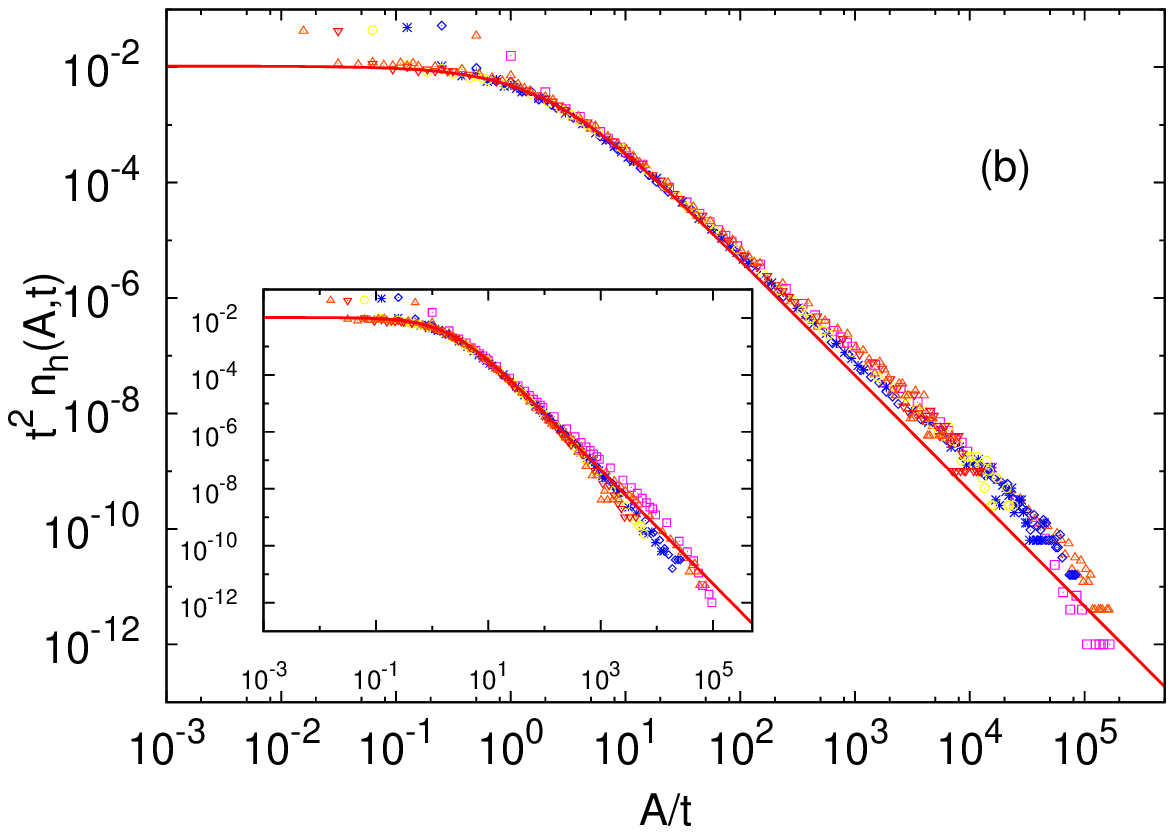}
\caption{(Colour online.)  Scaling plot of the number density of
  hull-enclosed areas in: (a) the experiment; (b) the $2d$IM with
  linear size $L=1280$ and periodic boundary conditions evolving with
  non-conserved order parameter at $T=0$. In the latter the
  measurements are done on a box with linear size $\ell=1000$.  The lines
  are the prediction in Eq.~(\ref{eq.analytic-nh}).  In the $2d$IM
  case we exclude the spanning clusters from the statistics. In the
  insets we exclude all domains that touch the border while in the
  main panels we include them in the statistics.}
\label{fig:exp-2dIM}
\end{figure}

We counted the number of hull-enclosed areas in $[A,A+dA]$ to
construct $n_h(A,t)$.  It is important to note that the experimental data are
taken using a {\it finite} imaging window and, in each sample and at
each measuring time, many domains touch the boundaries (in contrast to
the numerical simulations in \cite{us-PRL,us-PRE,us-EPL} in which we
used periodic boundary conditions). The zero-chirality constraint is
not obeyed exactly, both because the image is just a subset of the
whole sample and also because of the thresholding operation.

\begin{figure}[h]
\includegraphics[width=6cm]{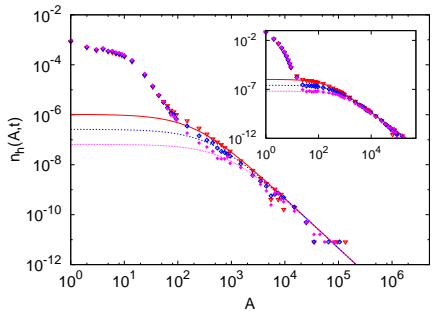}
\caption{(Colour online.)  The number density of hull-enclosed areas
  for the liquid crystal sample at $t=100,
  200, 400$s, excluding from the analysis the areas that touch the box
  border.  Inset: $n_h$ for the $2d$IM with $10\%$ randomly flipped
  spins at measurement (see the text for an explanation).  In both
  cases the lines are the theoretical prediction
  (\ref{eq.analytic-nh}) with  $c_h=1/8\pi\sqrt{3}$ and $\lambda_h=2.1$.}
\label{fig:pdf-exp}
\end{figure}

In Fig.~\ref{fig:exp-2dIM}(a) we show the hull-enclosed area
distribution in the liquid crystal at three different times. In the
main panel we included in the statistics the chopped areas that touch
the border of the image. The upward deviation of the data with respect
to the asymptotic power law $A^{-2}$ is due to the finite image
size. Indeed, domains that touch the border are actually larger but
get chopped and contribute to bins of smaller $A$'s and this induces a 
bias in the data. The inset displays $n_h$
removing from the statistics the areas that touch the border. The same
anomaly appears in the $2d$IM if one uses a finite imaging box within
the bulk. To show this we simulated a system with $L=1280$ and
periodic boundary conditions and we measured $n_h$ in a finite square
window with linear size $\ell=1000$ using $100$ independent samples.
In Fig.~\ref{fig:exp-2dIM}(b) we show two sets of data for the $2d$IM;
in both cases we exclude the spanning cluster over the full system
size.  One set of data includes areas touching the border and lies
above the theoretical curve. In the other set we eliminated
these areas from the statistics and the datapoints fall on the
analytic curve recovering the $A^{-2}$ tail.

The data in Fig.~\ref{fig:pdf-exp} do not show any noticeable
time-dependence at either small or large $A$. In the small area limit
the time-independence can be traced back to the lack of
time-dependence in the correlation function at
distances $r\stackrel{<}{\sim} 5$ (which corresponds to $A \simeq \pi r^2
\stackrel{<}{\sim} 80$), roughly the scale of the spatial graininess
(see Fig.~\ref{fig.correlation}). In the large area limit the
time-dependence naturally disappears; structures with $A\gg R^2(t)$
are basically the ones already present in the initial condition and
have not had time to evolve yet. In between these two limits the
curves show a shoulder with a systematic time-dependence that is the
most relevant part of our experimental data and it is very well
described by the analytic prediction (\ref{eq.analytic-nh}) shown with
solid lines.  To conclude we show that the random spins introduced by
the measuring method are not only responsible for the
time-independence of $n_h$ at small areas but also for the excess
weight of the distribution in this region.  In the inset to
Fig.~\ref{fig:pdf-exp} we show the hull-enclosed area distribution in
the $2d$IM where we introduced $10\%$ random spins at each measuring
time. There is indeed a strong similarity with the experimental data in the
main panel that could even be improved by choosing to flip spins in a
fine-tuned correlated manner.

In summary, our experimental results for the hull-enclosed area distribution in the coarsening
dynamics of the liquid crystal are in very good agreement
with the exact analytic prediction for $2d$ non-conserved scalar order
parameter dynamics presented in~\cite{us-PRL,us-PRE}.

JJA is partially supported by the Brazilian agencies CNPq, CAPES and FAPERGS,
LFC is a member of IUF. JMP, IA and ICP thank the Basque Government,
the MEC of Spain, respectively, for support.
This work was partially supported by the
CICYT-FEDER of Spain-EU (Project No. MAT2006-13571)

\end{document}